\documentclass[english]{paper}
\usepackage[T1]{fontenc}
\usepackage[latin9]{inputenc}
\usepackage{float}
\usepackage{textcomp}
\usepackage{amsmath}
\usepackage{amssymb}
\usepackage{graphicx}

\makeatletter

\providecommand{\tabularnewline}{\\}

\newcommand{\lyxaddress}[1]{
\par {\raggedright #1
\vspace{1.4em}
\noindent\par}
}

\makeatother

\usepackage{babel}
\begin{document}

\title{Risk Mathematics and Quantum Games on Quantum Risk Structures - A
Nuclear War Scenario Game}

\author{Carlos Pedro Gonçalves}

\institution{Instituto Superior de Ciências Sociais e Políticas (ISCSP) - Technical
University of Lisbon. }

\maketitle

\lyxaddress{E-mail: cgoncalves@iscsp.utl.pt}
\begin{abstract}
Quantum game theory is combined with risk mathematics' formalism to
provide an approach to evolutionary scenario analysis. The formalism
is addressed in its general form and is then applied to an extreme
risks modelling case, to model a coevolving dynamical web of systemic
situations representing the evolution of the regional tensions between
two countries with nuclear weapons. The model's results are addressed
regarding the potential for regional nuclear conflict to take place,
and how evolutionary scenario analysis may contribute to nuclear war
threat assessment and dynamical risk analysis. A final discussion
is provided in what regards risk mathematics based on the evolutionary
approach to risk assessement resulting from the combination of quantum
game theory, morphic web representations and scenario analysis.\end{abstract}

\section{Introduction}

A key problem in risk assessment is the combination of scenario analysis
tools with game theory to be applied to evolving unstable conditions
and complex systems' dynamics \cite{key-2,key-11,key-14}, quantum
game theory has proven a successful tool in dealing with such dynamics,
in particular, in what regards financial turbulence modelling \cite{key-12}.

In the present work, quantum game theory is combined with risk mathematics'
formalism to address evolutionary scenario dynamics. In \emph{section
2.}, a review is provided regarding the empirical effectiveness of
quantum game theory and its conceptual and systemic foundations, integrating
the current work within the broader background of quantum game theory
and quantum computation.

In \emph{section 3.}, the formalism for the integration between risk
mathematics and quantum game theory is provided, within a mathematical
foundation that combines formal structures coming from risk mathematics
with quantum computation theory (\emph{subsection 3.1}) and modal
systemics%
\footnote{By modal systemics we mean an approach to modal logic based upon a
systems science's ontological foundation, sharing the same conceptual
basis of risk mathematics, as developed in \emph{subsections 3.1}
and \emph{3.2}.%
} (\emph{subsection 3.2}), to address quantum games on \emph{quantum
risk structures}.

The formalism is then exemplified, in \emph{section 4.}, through a
geopolitical dynamics model between two countries with nuclear weapons,
providing for an approach to nuclear war threat assessment. General
conclusions are drawn, in \emph{section 5.}, regarding evolutionary
scenario analysis and risk mathematics' theory and applications.

\section{Quantum Game Theory}

Quantum game theory has been shown to hold with a good empirical matching
in capturing social learning dynamics and in economic and financial
decisional contexts \cite{key-13}. A particularly effective application
of quantum game theory regards risk contexts which include the analysis
of financial crises \cite{key-13} and financial turbulence modelling
\cite{key-12}. The good empirical matching of quantum game theory
within quantum econophysics opens up the matter of other applications
of quantum game theory, in particular, in areas such as political
science and strategic studies.

Quantum game theory fits well in non-equilibrium dynamics of complex
adaptive systems, since the probability structures may change with
the system's path-dependent quantum computation and result from the
coevolving relational structure that links system with its environment.

In this way, one may have dynamical instability rather than stable
fixed point strategies, and coevolving game conditions rather than
a fixed framework to which players must conform. The rules of the
game become more fluid and conditions may change suddenly and unexpectedly.
This is possible due to the underlying framework of path-dependent
quantum computation, which opens up the possibility of chaotic and
stochastic quantum logical gate updating (an approach that has been
followed in financial turbulence modelling \cite{key-12}) and opens
up the way to coevolving game conditions where the rules, themselves,
are permanently enacted towards an adaptive effectiveness of game
players.

In a systemic foundational framework, quantum game theory incorporates
quantum computation bringing it to an adaptive setting \cite{key-23,key-5,key-25}
which expands an argument laid out by Everett \cite{key-4} for a
fundamental systemic accounting of the systems' dynamics in the universe,
an argument that was effectively expanded by Gell-Mann and Hartle
in their reflection on \emph{quantum mechanics in light of quantum
cosmology} \cite{key-7}. In this reflection, Gell-Mann and Hartle
consider the strong dependence upon the quantum dynamical substratum
of the regularities exploited by what they call the \emph{environmental
sciences} such as astronomy, geology and biology, which are ultimately
traceable to the universe's initial condition, involving correlations
that stem from that initial condition.

Everett's legacy, either in the form of the many worlds interpretation
\cite{key-3} or in the form of the possible histories' interpretations%
\footnote{It has also been called \emph{decoherent histories interpretation}
\cite{key-7,key-8}, as well as \emph{consistent histories interpretation}
\cite{key-19}, we call it \emph{possible histories} due to the linkages
to modal systemics, since, in this interpretation, one assumes the
\emph{world in act} (our universe) and addresses the possible alternatives
for that \emph{world in act}, applying the formalism of quantum mechanics
\cite{key-7,key-8,key-19,key-20}%
} \cite{key-7,key-19}, offers a quantum physical basis for addressing
a quantum game theory applied to an understanding of complex adaptive
systems. A matter that is addressed within a quantum computational
perspective on the universe, defended by Deutsch \cite{key-3} and
by Lloyd \cite{key-20}, and within a quantum theoretical approach
to the adaptive cognition and information processing of complex adaptive
systems, defended by Gell-Mann and Hartle \cite{key-7} through the
notion of \emph{information gathering and utilization systems} (IGUSes),
which expands Everett's computational framework for the observer that
is described by Everett \cite{key-4} in terms of a computational
system equiped with a physical interface with the universe and memory
working that, in Everett's theory, make the observer necessarily entangled
with other systems in the universe.

In the current work we also take another step in addressing complex
adaptive systems' dynamics quantum theoretically, through the combination
of quantum game theory with risk mathematics. This leads us to an
evolutionary interpretation of the quantum state vector that links
the quantum probability measures to a \emph{fitness measure} defined
in terms of a relative frequency of alternative possibile configurations
of the universe/\emph{world in act}. Albeit following Everett's framework,
the mathematical results presented in the next section provide for
what can be called a {}``weak'' many worlds interpretation, in the
sense that it works by reference to the possibilites from a \emph{world
in act}.

What we call here the {}``strong'' many worlds interpretation assumes
parallel alternatives all occurring in actualized branches, while,
for the {}``weak'' interpretation, the \emph{world in act}, our
universe, addressed quantum theoretically within the framework of
complex quantum systems science, is considered to be in a permanent
(quantum) computation, with the final resulting actualized configuration
being systemically selected through an evolutionary process out of
a systemic evaluation of alternatives, thus, out of the possible alternatives,
one takes place in act, which allows the theory to approach the universe
as an evolving complex quantum system, an approach that comes from
both Deutsch and Lloyd \cite{key-3,key-20}, as well as Gell-Mann
and Hartle \cite{key-7,key-8} being rooted in Everett's work \cite{key-4}.

The {}``strong'' many worlds interpretation has been used in quantum
game theory \cite{key-25} as an effective way to consider how different
decisions may lead to parallel realities in a relational universe.
Within an evolutionary framework, Everett's notion of measure \cite{key-4},
worked from the {}``strong'' many worlds interpretation of quantum
mechanics \cite{key-3}, can be addressed such that the occurrence
and non-occurrence of the events leads to a statistical distribution
over an ensemble of parallel worlds that reflects the systemic sustainability
of each alternative in what can be considered a relative frequency
\emph{fitness measure}, such that, like a genetic type's relative
frequency in a population constitutes a measure of that genetic type\textquoteright{}s
\emph{fitness} (relative frequency measure of \emph{fitness}), so
one may assume to take place in the {}``strong'' many worlds interpretation,
where the branches of a quantum state are interpreted as leading to
a measure of each alternatives' representativeness in the ensemble
of parallel realities, thus alternatives with higher \emph{fitness}
in this ensemble become the ones that are more frequently observed
and, simultaneously, the more likely paths for the systems.

A similar approach is followed here, only that the selection is for
a \emph{single world in act} between an ensemble of possible configurations
of that \emph{world in act}, with the \emph{fitness measure} holding
for such an ensemble.

One of the reasons for our choice to work with the {}``weak'' many
worlds interpretation is the problem that the {}``weak'' interpretation
might have to be invoked in addressing the {}``strong'' interpretation,
in particular, to address how each configuration takes place in \emph{parallel
worlds in act}, leading to a specific statistical distribution of
actualized alternatives, so that the structure of the {}``weak''
interpretation might have to be assumed for each \emph{parallel world
in act}, this would lead to a doubling of the formalism, becoming
more parsimonious to assume and work with the {}``weak'' interpretation
right from the start.

However, this does not mean that a physical negation of parallel universes
is being proposed, indeed, as stated, the {}``weak'' interpretation
is compatible with both the {}``strong'' interpretation, that assumes
parallel \emph{worlds in act}, as well as with the view of those that
choose not to assume \emph{parallel worlds in act}. Since Physics
is divided on this matter, we do not take a stand here, because taking
a stand would lead us outside the scope of the present work.

The compatibility of the {}``weak'' interpretation with the {}``strong'',
as explained in the previous paragraphs, also makes the current work
compatible with the other works in quantum game theory that assume
the {}``strong'' interpretation \cite{key-25}, however, by assuming
the {}``weak'' interpretation we are able to integrate in a single
quantum game theory-based formal system, the three frameworks: risk
mathematics, quantum game theory and a modal systemics.

\section{Risk Mathematics, Quantum Games and Quantum Risk Structures}

Risk mathematics takes risk, itself, as its object of research and
addresses it systemically from its root in the Medieval Latin term
\emph{resicum}, which synthesized, in the context of \emph{Maritime
Law}, the three notions of \emph{periculum} (peril, threat), \emph{fortuna}
(fortune, luck, destiny) and uncertainty: to play for one's destiny
in situations where there are threats and opportunities, with an outcome
still open or unknown to the player \cite{key-10,key-11,key-21}.
Thus, when one addresses a risk situation (for instance, a nuclear
war between two countries), one is addressing a systemic configuration
that is ontologically contingent, and epistemologically non-determined,
that is, it may or may not take place, so that it is not an impossibility,
and there is uncertainty with regards to its taking place.

The recognition of a risk situation, therefore, demands the recognition
of the presence of the systemic ground for a threat. Thus, for instance,
the presence of a gun in a place opens up immediately the possibility
of that gun being used in a myriad of risk situations. The interconnectedness
of risk situations has led authors to address risk scenarios in terms
of network structures \cite{key-10,key-11,key-14}, considering the
influence webs between different risk situations, these network structures
can be further formalized in terms of the notion of \emph{morphic
webs} \cite{key-9,key-10,key-11}. 

Formally, within mathematics, a \emph{morphic web} is a weaker structure
than a category, being defined as a mathematical structure $\mathcal{W}$
of objects $A,B,C,...$ and morphisms, between objects $A\overset{f}{\longrightarrow}B$,
with the identity morphism connecting each object to itself $A\overset{id_{A}}{\longrightarrow}A$.
A mathematical category is a \emph{morphic web} with further axioms
regarding composition of morphisms%
\footnote{In particular: closure, associativity and identity laws ($id_{A}\circ f=f\circ d_{B}=f$,
for $A\overset{f}{\longrightarrow}B$) with respect to the composition
operation.%
}, however, not all\emph{ morphic webs} are categories \cite{key-9,key-11}.

A \emph{morphism} synthesizes a directional relation from an origin
object to a target object, expressing a systemic motion from the origin
to the target with relational fundament $f$ \cite{key-9}.

\emph{Morphic webs} can be applied as effective tools in a systemic
analysis of risk, in particular in what regards scenario analysis
built from morphic webs of systemic situations \cite{key-11}, defined
as morphic webs $\mathcal{W}$ whose object collection, denoted by
$\textrm{ob}\left(\mathcal{W}\right)$, is a collection of systemic
situations and whose morphism collection, denoted by $\textrm{morph}\left(\mathcal{W}\right)$,
is composed of morphisms of the kind $A\overset{f_{AB}}{\longrightarrow}B$
which are interpreted as $A$ being a \emph{systemic source} of $B$,
and $f_{AB}$ is a number representing the coupling strength from
$A$ to $B$. Scenarios can be built by addressing the different combinations
of ocurrences and influences, as explained in \cite{key-11}.

However, to deal with evolutionary dynamics of scenarios, rather than
a static scenario enumeration, one needs to expand the formal basis
of \emph{morphic webs}, so that one may address, in evolutionary terms,
the occurrence of events and the dynamical linkages influenced by
the morphic structure of the systemic situations' web. It turns out
that quantum game theory allows one to expand the \emph{morphic webs}'
formal basis and to address such a dynamical evolutionary setting,
as we now show.

\subsection{Quantum risk structures}

To build a quantum game theoretical approach to evolutionary scenario
analysis, within the framework of risk mathematics, we need to consider,
first, a morphic web of systemic situations $\mathcal{W}$ and introduce
the Hilbert space $\mathcal{H}$ such that $\mathcal{H}:=\bigotimes_{i=1}^{n}\mathcal{H}_{i}$,
where $\mathcal{H}_{i}$ is the Hilbert space associated with the
\emph{i}-th systemic situation%
\footnote{Formally, the \emph{i}-\emph{th} object.%
} of $\mathcal{W}$, for $n=\#\textrm{ob}\left(\mathcal{W}\right)$
(the number of systemic situations in $\mathcal{W}$), we also assume
that each $\mathcal{H}_{i}$ is spanned by the basis $\left\{ \left|0\right\rangle ,\left|1\right\rangle \right\} $,
where $\left|0\right\rangle $ encodes the case in which the $i$-th
systemic situation does not occur and $\left|1\right\rangle $ encodes
the case in which the $i$-th systemic situation occurs.

Introducing, for each $\mathcal{H}_{i}$, the basis projectors $\hat{P}_{s}^{i}=\left|s\right\rangle \left\langle s\right|$,
with $s=0,1$, we build the full projector space $\mathcal{P}:=\left\{ \hat{P}_{\mathbf{s}}=\bigotimes_{i=1}^{n}P_{s_{i}}^{i}:s_{i}=0,1\right\} $,
with $\mathbf{s}=\left(s_{1},s_{2},...,s_{n}\right)$, so that each
$\hat{P}_{\mathbf{s}}$ projects onto a basis vector of $\mathcal{H}$
and corresponds to a scenario of occurrences over $\mathcal{W}$.
By scenario we understand a description of a course of events or situations
that may take place (are possible), thus a scenario is a complete
description of systemic occurrences in a \emph{world configuration}
that is possible for the \emph{world in act} (the \emph{universe}),
each scenario is, in this case, encoded by the syntax of a binary
quantum basis alphabet with the semantics provided by $\mathcal{W}$
which gives us the account of the scenario\emph{ }itself.

The general \emph{ket} vector of $\mathcal{H}$ expands as follows:
\begin{equation}
\left|\Psi\right\rangle =\sum_{\mathbf{s}}\hat{P}_{\mathbf{s}}\left|\Psi\right\rangle =\sum_{\mathbf{s}}\psi\left(\mathbf{s}\right)\left|\mathbf{s}\right\rangle 
\end{equation}
with $\psi\left(\mathbf{s}\right)=\left\langle \mathbf{s}|\Psi\right\rangle $.
Thus, each branch of $\left|\Psi\right\rangle $ corresponds to an
alternative configuration of the universe that can take place in act,
with an amplitude $\psi\left(\mathbf{s}\right)$, different scenarios
that can take place in act correspond to different possible configurations
of the universe (different \emph{world configurations}), the universe
would have different information contents, for instance, if the South
had won the American Civil War.

One must take care in addressing the formal syntax and the semantics
underlying Eq.(1), in effect we have \emph{qubitized} the information
contents of the \emph{possible world configurations} with respect
to the description of whether or not each systemic situation in $\mathcal{W}$
occurs, this description constitutes the semantics for the syntax,
but it also allows us to formally assign a \emph{n}-qubit state as
per Eq.(1), with each branch from the state vector $\left|\Psi\right\rangle $
synthesizing a complete pattern of occurrences. This \emph{qubitization}
can only be done due to the fact that $\left|\Psi\right\rangle $
is an account of scenarios taking place in possible configurations
for the universe, as stated in the previous paragraph.

The vector expansion in $\left|\Psi\right\rangle $, thus, shows a
\emph{n-}qubit state\emph{ }describing the alternative\emph{ }scenarios,
the semantics for this quantum computational syntax is provided by
the systemic situations' \emph{morphic web} $\mathcal{W}$, which
gives us how each alternative $\hat{P}_{\mathbf{s}}$ can be interpreted,
this allows us to introduce a complete (quantum) logical semantic
structure that can be defined as:
\begin{equation}
\mathfrak{A}\left(\left|\Psi\right\rangle \right):=\left(\mathcal{W},\mathcal{H},\mathcal{P},\left|\Psi\right\rangle \right)
\end{equation}
This last structure, introduced within the (quantum) model logical
semantics, corresponds to what we call a \emph{quantum risk structure}.
A \emph{quantum risk structure} is similar in many respects to Gell-Mann
and Hartle's proposal of addressing the quantum state of the universe
in terms of an exhaustive set of projectors that project for yes or
no questions \cite{key-7}, an approach that comes from Everett's
line of argument \cite{key-4,key-7}, linking the systems within the
universe as systems in the universe and therefore within a quantum
cosmological setting, working from the universe's quantum state with
exhaustive sets of alternatives \cite{key-7,key-8,key-19}, which
provides for a physical and systemic framework for general risk science.

To address risk, however, it becomes necessary to systemically interpret
the quantum amplitudes $\psi\left(\mathbf{s}\right)$, so that we
need to have a formalism that allows us to address the systemic possibility
of occurrences of each alternative scenario connecting the amplitudes
to this systemic possibility and, in turn, we need to extract from
$\left|\Psi\right\rangle $ information that allows us to introduce
probabilities for how different scenarios can take place in act. This
can only be done by addressing the notion of possibility within a
modal systemics.

\subsection{Modal systemics and quantum probabilties}

Modal logics can be built from formal languages to address \emph{ontological
modalities}, including, in particular, necessity and possibility \cite{key-1,key-6},
we need to address a specific formal structure that is simultaneously
consistent with quantum game theory and with the systemic framework
of the \emph{quantum risk structures} introduced in the previous subsection%
\footnote{Within formal mathematics, the construction of formal structures for
concrete systems is sometimes non-univocal, in the sense that different
paths to formalization can be built. In interdisciplinary applications,
the choice becomes one of effectiveness and consistency with the different
sciences involved. In the present case, we have systems science, risk
mathematics and quantum game theory.%
}. A solution is to introduce a \emph{quantum modal structure} which
includes the \emph{quantum risk structure} $\mathfrak{A}\left(\left|\Psi\right\rangle \right)$,
defined as:
\begin{equation}
\mathfrak{W}\left(\left|\Psi\right\rangle \right):=\left[\mathbf{W},\mathfrak{A}\left(\left|\Psi\right\rangle \right)\right]
\end{equation}
where $\mathbf{W}$ is a morphic web comprised of one object $w_{0}$
representing the \emph{world in act} and where possibility is addressed
from the \emph{world in act} in terms of possible configurations of
that world or \emph{possible world configurations} that comprise the
morphic structure of possibilities $\mathbf{W}$ and are formally
built as:
\begin{equation}
w_{0}\overset{\mathbf{c}}{\longrightarrow}w_{0}
\end{equation}
where the fundament $\mathbf{c}$ corresponds to a specific \emph{possible
world configuration} of the \emph{world in act}.

Given the \emph{morphic web} of systemic situations $\mathcal{W}$,
a morphic structure of possibilities $\mathbf{W}$ consistent with
$\mathcal{W}$ is such that for each morphism $w_{0}\overset{\mathbf{c}}{\longrightarrow}w_{0}$,
there is one, and only one, scenario of occurrences $\mathbf{s}$
of the systemic situations of $\mathcal{W}$ that occurs in $\mathbf{c}$,
we, thus, write $\mathbf{s}\in\mathbf{c}$.

While, for each configuration, there can only be a single occurring
scenario, for different configurations the same scenario may take
place. Since the projection of $\left|\Psi\right\rangle $ over each
basis element $\left|\mathbf{s}\right\rangle $ provides for quantum
amplitudes $\psi\left(\mathbf{s}\right)\left(=\left\langle s\left|\hat{P}_{\mathbf{s}}\right|\Psi\right\rangle \right)$
over each alternative scenario $\mathbf{s}$, these amplitudes must
somehow be linked to the distribution of occurrences of $\mathbf{s}$
over $\mathbf{W}$. We, therefore, need to find, within quantum theory,
a conceptual scheme to extract information from $\left|\Psi\right\rangle $
on the distribution of configurations of $\mathbf{W}$ with respect
to the occurrences of the scenarios that come from $\mathcal{W}$
and we need to connect such a distribution to a game theoretical evolutionary
interpretation. Such a conceptual scheme can be introduced through
the \emph{decoherence functional }\cite{key-7,key-8,key-19,key-24}
for the scenarios, defined as:
\begin{equation}
D\left(\mathbf{s},\mathbf{s}'\right):=\left\langle \Psi\left|\hat{P}_{\mathbf{s}'}^{\dagger}\hat{P}_{\mathbf{s}}\right|\Psi\right\rangle 
\end{equation}
which satisfies:
\begin{equation}
D\left(\mathbf{s},\mathbf{s}'\right)=\left|\psi\left(\mathbf{s}\right)\right|^{2}\delta_{\mathbf{s},\mathbf{s}'}\geq0
\end{equation}
\begin{equation}
\sum_{\mathbf{s},\mathbf{s}'}D\left(\mathbf{s},\mathbf{s}'\right)=\sum_{\mathbf{s}}\left|\psi\left(\mathbf{s}\right)\right|^{2}=1
\end{equation}
therefore, $D\left(\mathbf{s},\mathbf{s}'\right)$ behaves like an
additive measure over the alternative scenarios. A further assumption
is then made that, for $\mathbf{W}$, the proportion of morphisms
$w_{0}\overset{\mathbf{c}}{\longrightarrow}w_{0}$ in which $\mathbf{s}$
occurs is equal to $D\left(\mathbf{s},\mathbf{s}\right)=\left|\psi\left(\mathbf{s}\right)\right|^{2}$,
within risk mathematics, this proportion is linked to the systemic
tendency for $\mathbf{s}$ to take place in the \emph{world in act},
so that we can measure from $\left|\psi\left(\mathbf{s}\right)\right|^{2}$
the probability of the scenario to take place in the \emph{world in
act} $w_{0}$.

Simultaneously, we can also provide for an evolutionary game theoretical
interpretation of $\left|\psi\left(\mathbf{s}\right)\right|^{2}$,
since the relative frequency occurrences of $\mathbf{s}$ in the morphisms
of $\mathbf{W}$ can be read as a measure of systemic sustainability
underlying the occurrence of $\mathbf{s}$, interpretable as a \emph{fitness}
associated with each branch $\hat{P}_{\mathbf{s}}$ of the quantum
state $\left|\Psi\right\rangle $. Applying Eq.(5) we can obtain a
\emph{fitness density operator} akin to evolutionary game theory's
version of a \emph{fitness} matrix:
\begin{equation}
\hat{\mathfrak{D}}\left(\left|\Psi\right\rangle \right):=\sum_{\mathbf{s'},\mathbf{s}}\left\langle \Psi\left|\hat{P}_{\mathbf{s}'}^{\dagger}\hat{P}_{\mathbf{s}}\right|\Psi\right\rangle \left|\mathbf{s}\right\rangle \left\langle \mathbf{s'}\right|
\end{equation}
from Eq.(6) it follows immediately that this \emph{fitness density
operator} is diagonal and given by:
\begin{equation}
\hat{\mathfrak{D}}\left(\left|\Psi\right\rangle \right)=\sum_{\mathbf{s}}\left|\psi\left(\mathbf{s}\right)\right|^{2}\hat{P}_{\mathbf{s}}
\end{equation}
from this diagonal form, in a game theoretical sense, one can state
that the scenarios are \emph{mixing} on the quantum strategy $\left|\Psi\right\rangle $,
in a quantum theoretical sense, one can state that the alternatives
$\hat{P}_{\mathbf{s}}$ decohere \cite{key-7,key-19,key-24}. In the
quantum game theoretical framework, $\hat{\mathfrak{D}}\left(\left|\Psi\right\rangle \right)$
has an evolutionary interpretation such that the selection of each
alternative $\mathbf{s}$ is systemically grounded in the \emph{fitness}
or \emph{systemic sustainability} for $\mathbf{s}$, leading to a
probability measure of $\mathbf{s}$ taking place in the \emph{world
in act}.

For such a game theoretical framework, there can be no collapse of
$\hat{\mathfrak{D}}\left(\left|\Psi\right\rangle \right)$ (nor of
$\left|\Psi\right\rangle $), since the possibilities always remain
as possibilities independently of what takes place in act (which is
also necessarily possible) \cite{key-21}, so that the \emph{morphic
web} $\mathbf{W}$ always holds with its statistical configuration.
Indeed, a potency is actualized by the act that determines it, one
must not, however, mix the notion of potency with the notion possibility:
possible is what can be, a potency (\emph{dynamis}) is towards the
act (\emph{energeia}), these are different notions tracing back to
Aristotle's thinking, in which modal logics find their ground \cite{key-1,key-6,key-21}.

Under the current modal framework, the probability for each alternative
to take place in act is numerically coincident with the \emph{fitness},
measuring how the selection may take place in act, which means the
probability with which an alternative scenario $\mathbf{s}$ is actualized
is taken as numerically coincident with the relative frequency $\left|\psi\left(\mathbf{s}\right)\right|^{2}$
of possible configurations of the world in act in which $\mathbf{s}$
occurs, after the actualization, on the other hand, the scenario $\mathbf{s}$
either took place in act or not, such that the probability is either
0 or 1, the \emph{fitness} density $\hat{\mathfrak{D}}\left(\left|\Psi\right\rangle \right)$
as well as the quantum state $\left|\Psi\right\rangle $ are still,
however, necessarily, unchanged, being related to the sustainability
field associated with the systemic situation, leading to the ensemble
distribution of possibilities for the \emph{world in act}.

It is important to notice that the probability only holds if the \emph{fitness}
levels $\left|\psi\left(\mathbf{s}\right)\right|^{2}$ remain unchanged,
otherwise, if a change in the \emph{fitness} of each alternative occurs
at the very last moment of actualization, the probability changes
and, thus, the system can change direction in its choice. Changes
in the \emph{fitness density} results from a quantum computation rule
which is akin to a \emph{quantum replicator dynamics} over possible
configurations of the \emph{world in act}, such that a quantum unitary
operator encodes the system's evolutionary computation. Indeed, the
modal structure $\mathfrak{W}\left(\left|\Psi\right\rangle \right)$
changes for quantum computations that transform the state $\left|\Psi\right\rangle $
and, thus, simultaneously, transform the configuration of the \emph{morphic
web} $\mathbf{W}$ with respect to the statistical distribution of
\emph{possible world configurations} containing each scenario.

For a quantum game divided in rounds, indexing the rounds by $t=1,2,...,$
with a unitary quantum computing gate holding at each round, we obtain:
\begin{equation}
\left|\Psi(t)\right\rangle =\hat{U}(t)\left|\Psi(t-1)\right\rangle 
\end{equation}
the unitary state transition changes the quantum game state in accordance
with the nature of the systemic situations and the systems involved
(in particular, it must reflect the morphic connections of $\mathcal{W}$).
It is important to notice that $t$ is a game round index, not a temporal
physical clock frame, it just counts the number of quantum computing
steps as per Eq.(10), thus, it matches the counting of quantum game
rounds. In this case, there is a corresponding round indexing of the
morphic web $\mathbf{W}(t)$ matching the indexing of $\left|\Psi(t)\right\rangle $,
so that we have the quantum game round-dependent structure $\mathfrak{W}\left(\left|\Psi(t)\right\rangle \right)=\left[\mathbf{W}(t),\mathfrak{A}\left(\left|\Psi(t)\right\rangle \right)\right]$.

The quantum game conditions can be evolve from round to round so that
the structure of unitary gates $\hat{U}(t)$ can show a path-dependence,
this path-dependence can be upon the sequence of the previous unitary
gates {\small $\hat{U}(0),...,\hat{U}(t-1)$}, but it can also depend
upon actualized alternatives. It is important to notice that the possible
histories interpretation has shown that actualization is not synonymous
with breakdown in unitary state transition from an initial condition,
rather, the work done within decoherence theory has shown that unitary
state transition can take place with an actualized history for the
system%
\footnote{A point made explicit for instance in Gell-Mann and Hartle's recent
work on an \emph{actualized fine-grained history} present simultaneously
with unitary state transition \cite{key-8}%
} \cite{key-7,key-8,key-20}.

The fact that several decoherence theory proposals follow and indeed
expand on Everett's approach \cite{key-16,key-17,key-18,key-19} shows
that the so-called breakdown in unitary state transition, with loss
of quantum interference, takes place either through entanglement with
a description in terms of local degrees of freedom monitoring, tracing
out the correlated environment \cite{key-16,key-18}, or due to coarse-graining
\cite{key-7,key-19} or even a mixture of both \cite{key-19}. In
the first case, it becomes necessary to re-express the local state
transition in terms of a quantum map reflecting the global unitary
state transition with local loss of information due to interaction
with a local environment \cite{key-17}. Hawking radiation is also
another possible source of unitary state transition breakdown, even
though Hawking has recently seemed to back out from this view on information
loss \cite{key-15}. However, in no instance is there an unequivocal
association between breakdown of unitarity and actualization, there
is no physical causality link that can be made mathematically between
the two, except by assuming it to be there as a postulate which is
extra-theoretical \cite{key-18}.

Indeed, Everettian decoherence theory has shown us that we can have
an actualized history of the world (or worlds in the case of the {}``strong''
many worlds interpretation), with sequences of unitary state transitions
without breakdown of quantum interference, the breakdown of quantum
interference can be stated to be due to a tracing out of entangled
states, or to coarse-graining, but this is an observer-related ignorance
effect. In no way does the mathematical formalism support consistently
the proposal that actualization is linked to breakdown of quantum
interference effects due to observer-related local description tracing
out the environment. Such a proposal is a meta-formalism postulate
that is inconsistent with the notion of actualization, introducing
to a conceptual confusion between possibility, potency and probability
which are, indeed, three distinct notions within Philosophy and Mathematics
as stated above. If one works with a notion of actualization, one
cannot properly assume that actualization takes place due to tracing
out degrees of freedom of an entangled system or due to coarse-graining,
and physical experiments have shown this to be true \cite{key-16,key-20}.

Having addressed the main formalism and conceptual substratum we now
provide an example of the current formalism applied to a nuclear war
threat assessment problem.

\section{A Nuclear War Quantum Game Model}

Let us consider a geopolitical game between two coutries, labelled
$A$ and $B$. The first step in building the game is to draw the
morphic web of systemic situations, we consider the following eight
situations, where the order of presentation and labelling has been
arbitrarily assigned:
\begin{itemize}
\item $S_{0}:$ growing political tensions between the two countries linked
to conflicts of interest, possible diplomatic incidents, economic
and political competition with possible military fallout.
\item $S_{1}$: political turmoil in the two countries, which includes internal
and external political turmoil involving the two countries' tension
issues, it can lead to both demonstrations and actions of supporters
of each country in the streets, and political instability due to political
statements and actions undertaken by both countries. 
\item $S_{2}$: failure of both countries to negotiate an aggreement on
key issues, which may prevent the countries from communicating and
finding diplomatic solutions to their grievances. 
\item $S_{3}$: civil unrest, that is, when people take to the streets in
protest on both internal problems and problems involving both countries,
it also includes the possibility of social disorder and possible escalade
in street violence linked to key problems that divide the two countries. 
\item $S_{4}$: conflict breakout between the two countries involving conventional
non-nuclear arsenal.
\item $S_{5}$: nuclear threat escalation, with the threat of both countries
using their nuclear arsenal.
\item $S_{6}$: Nuclear war, this corresponds to both countries using nuclear
weapons on each other, triggering a nuclear war.
\item $S_{7}$: United Nations (UN) intervention/mediation, corresponding
to attempts of UN to influence both countries in finding common ground
and mediating political negotiations between both countries, but it
also entails possible deliberations and actions that can be taken
by the UN Security Council in the case of military conflict between
the two countries or of nuclear weapons being used.
\end{itemize}
The morphisms are the following:
\begin{itemize}
\item With origin $S_{0}$: $S_{0}\overset{f_{0j}}{\longrightarrow}S_{i}$,
with $i=1,2,3,4,6$;
\item With origin $S_{1}$: $S_{1}\overset{f_{1j}}{\longrightarrow}S_{i}$,
with $i=0,2,3$;
\item With origin $S_{2}$: $S_{2}\overset{f_{2j}}{\longrightarrow}S_{i}$,
with $i=0,1,3,4,7$;
\item With origin $S_{3}$: $S_{3}\overset{f_{3j}}{\longrightarrow}S_{i}$,
with $i=0,1,4,7$;
\item With origin $S_{4}$: $S_{4}\overset{f_{4j}}{\longrightarrow}S_{i}$,
with $i=0,1,2,5,7$;
\item With origin $S_{5}$: $S_{5}\overset{f_{5j}}{\longrightarrow}S_{j}$,
with $j=0,1,2,6,7$;
\item With origin $S_{6}$: $S_{6}\overset{f_{67}}{\longrightarrow}S_{7}$.
\end{itemize}
Each connection strength is set to one of three levels: low, medium
and high (with each level uniformly chosen to lie in the following
ranges: 0 to 0.1 (low connection strength); 0.1 to 0.4 (medium connection
strength); 0.4 to 0.9 (strong connection strength)).

Thus, given the links above, growing tensions between the two countries
are set at the morphic origin of the systemic situations: political
turmoil (medium connection strength); failure to negotiate (medium
connection strength); civil unrest (low connection strength), conflict
break out (low connection strength) and UN intervention/mediation
(medium connection strength).

Political turmoil, on the other hand, is linked, with high connection
strength, to: growing tensions; failure to negotiate and civil unrest.
Failure to negotiate is linked, also with high connection strength
to: growing tensions; political turmoil; civil unrest; conflict break
out and UN intervention/mediation.

Civil unrest is linked with medium connection strength to conflict
break out and with high connection strength to: growing tensions;
political turmoil and UN intervention/mediation.

Conflict break out, on the other hand, is linked with high connection
strength to growing tensions, political turmoil and failure to negotiate
and it is linked with medium connection strength to nuclear threat
escalation, in this case, the connection strength to UN intervention/mediation
is non-randomly set to 1 (maximum connection strength).

Nuclear threat escalation and nuclear war are also set with connection
strength of 1 to UN intervention/mediation, while nuclear threat escalation
is set with high connection strength to: growing tensions; political
turmoil; failure to negotiate and nuclear war. 

This defines the \emph{morphic web}%
\footnote{For concrete countries, the web might be expanded to include tension
triggering elements and other factors that are specific to the countries
under analysis, as it stands the current \emph{morphic web} constitutes
a general work basis for a quantum game adaptable to different regional
conflicts cases, being an example of how the formalism may be applied.%
}. Now, the round-dependent quantum unitary gate for the game is, in
the present case, set as:

\begin{equation}
\hat{U}(t)=\hat{U}_{+}(t)\hat{U}_{-}(t)
\end{equation}
\begin{equation}
\hat{U}_{+}(t)=\bigotimes_{i=0}^{7}\hat{U}_{i}\left(r_{i}(t)\right),\,\hat{U}_{-}(t)=\bigotimes_{i=0}^{7}\hat{U}_{i}\left(r_{i}(t-1)\right)^{\dagger}
\end{equation}
where $r_{i}(t-1)$ and $r_{i}(t):=F\left[r_{i}(t-1)\right]$ are
local evolutionary dynamical parameters and $F$ is a real-valued
map, implemented by the local unitary operators $\hat{U}_{i}$ which
are defined by the quantum logical gate:
\begin{equation}
\hat{U}_{i}\left(r\right)=\alpha_{i}(r)\left(\left|0\right\rangle \left\langle 0\right|-\left|1\right\rangle \left\langle 1\right|\right)+\beta_{i}(r)\left(\left|0\right\rangle \left\langle 1\right|+\left|1\right\rangle \left\langle 0\right|\right)
\end{equation}
\begin{equation}
\alpha_{i}(1-r)=\sqrt{1-r},\:\beta_{i}(r)=\sqrt{r}
\end{equation}
where $r$ is taken as a real number between $0$ and $1$ and, as
in the previous section's formalism, $\left|0\right\rangle $ encodes
the case in which the scenario $S_{i}$ does not take place and $\left|1\right\rangle $
encodes the case in which the scenario takes place. This leads to
a path-dependent quantum computation for each $S_{i}$, with the quantum
state for each round being given by:
\begin{equation}
\left|\Psi(t)\right\rangle =\bigotimes_{i=0}^{7}\left|\psi_{i}(t)\right\rangle =\bigotimes_{i=0}^{7}\hat{U}_{i}\left(r\right)\hat{U}_{i}\left(r_{i}(t-1)\right)^{\dagger}\left|\psi_{i}(t-1)\right\rangle 
\end{equation}
with, as per Eqs.(11) to (14), $r_{i}(t)=F\left[r_{i}(t-1)\right]$.

In the current quantum game, the map $F$ is an actualization-contingent
map, so that if $S_{7}$ (UN intervention/mediation) does not take
place in act, then $F$ is given by 
\begin{equation}
F\left(r_{i}(t-1)\right)=G\left(r_{i}(t-1)\right)
\end{equation}
where $G\left(r_{i}(t-1)\right)$ is a lattice coupled nonlinear map
to be specified shortly. On the other hand, if $S_{7}$ takes place
in act, then, $F$ is, for all systemic situations except $S_{7},$
given by:
\begin{equation}
F\left(r_{i}(t-1)\right)=\left(1-\theta_{UN}\right)G\left(r_{i}(t-1)\right)+\theta_{UN}\left(1-d_{UN}\right)G\left(r_{i}(t-1)\right)
\end{equation}
where the parameters $0\leq\theta_{UN}\leq1$ and $0\leq d_{UN}\leq1$
correspond, respectively, to the effectiveness of UN intervention/mediation
and $d_{UN}$ is the impact of UN intervention/mediation%
\footnote{If we were to consider here the {}``strong'' many worlds interpretation
one might assume a $\left|\Psi(0)\right\rangle $ as holding for each
parallel universe, then, for those universes/\emph{worlds in act}
in which the UN did not intervene, the state vector would evolve in
accordance with the unitary gate $\hat{U}\left(F\left(r_{i}(1)\right)\right)$,
with $F$ obeying Eq.(16), while, for those worlds in which UN did
intervene, the state vector would evolve in accordance with the unitary
state $\hat{U}\left(F\left(r_{i}(1)\right)\right)$, with $F$ obeying
Eq.(17), the intervention and the non-intervention would thus become
frozen events on each parallel universe holding from then on and leading
to a divergence from $\left|\Psi(0)\right\rangle $, with \emph{fitness
amplitudes} tracing back to the {}``ancestor'' state $\left|\Psi(0)\right\rangle $,
in this case, actualization leads to a form of mixed state for the
actualized multiverse description, since different worlds would be
described by different pure states, evolving from an initial pure
state that held for all, and one would have to account for this with
an ensemble-based multiverse description where the initial ensemble
is all in the same quantum state. This gives another perspective on
branching, and shows how, in a multiverse description, one might have
that unitary evolution path-dependent upon actualization would lead
to an effective mixedness rather than a reduced mixed density operator
due to a tracing out of a part of an entangled system's degrees of
freedom.%
}. For the UN intervention/mediation the dynamics is, in this case,
still given by Eq.(16). The nonlinear map $G(.)$, that appears in
both Eqs.(17) and (16), is a coupled nonlinear map with the following
structure:
\begin{equation}
G\left(r_{i}(t-1)\right)=\left(1-\varepsilon-\delta\right)M_{i}(t)+\varepsilon h_{i}(t)+\delta z_{i}(t)
\end{equation}
where $M_{i}(t)$ is the result of the logistic map dynamics upon
$\beta_{i}\left(r_{i}(t-1)\right)$ as follows%
\footnote{The logistic map expresses a growth limited by a carrying capacity
which is useful in capturing a nonlinear dynamics of political resources
that limit the decision to take a certain action. In the present case,
the noisy chaotic dynamics leads to a noisy chaotic update of each
qubit as per Eq.(15).%
}: 
\begin{equation}
M_{i}(t):=b\cdot\beta_{i}\left(r_{i}(t-1)\right)\left(1-\beta_{i}\left(r_{i}(t-1)\right)\right)
\end{equation}
and $h_{i}(t)$ is the nonlinear mean field quantity:
\begin{equation}
h_{i}(t):=\frac{1}{\sum_{j}f_{ji}}\sum_{j}f_{ji}M_{j}(t)\cdot M_{i}(t)
\end{equation}
where the sum is over all the systemic situations $S_{j}$ that are
at the morphic orginin of $S_{i}$ that is, all the systemic situations
that satisfy $S_{j}\overset{f_{ji}}{\longrightarrow}S_{i}$, weighted
by the respective morphic web couplings $f_{ji}$. The quantity $z_{i}(t)$
is a random uniform noise term between 0 and 1, thus, we have a noisy
nonlinear coupled map lattice, emerging from the quantum computing
dynamics due to the path-dependence and driving the quantum state
transition.

At beginning of the game we set the state:
\begin{equation}
\left|\Psi(0)\right\rangle =\sqrt{1-r_{0}(0)}\left|000...0\right\rangle +\sqrt{r_{0}\left(0\right)}\left|100...0\right\rangle 
\end{equation}
where $r_{0}(0)$ is a real number uniformly chosen between 0 and
0.01, this means that the whole interconnected risk dynamics is driven
by an initially small quantum amplitude for political tension between
the two countries, apart from random factors integrated in the uniform
noise coupling of Eq.(18). Thus, from the above equations, each unitary
transition leads to a proportion of $\left|\alpha_{i}\left(r_{i}(t)\right)\right|^{2}=r_{i}(t)$
of morphisms of $\mathbf{W}(t)$, where $S_{i}$ occurs and a proportion
of $1-r_{i}(t)$, where the event does not occur.

Taking into account Eq.(21), and from Eqs.(11) to (15) it follows
that, for each qubit $\left|\psi_{i}(t)\right\rangle $, we can writte
the difference equation:
\begin{equation}
\begin{aligned}\triangle\left|\psi_{i}(t)\right\rangle =\\
 & =\left\{ \alpha_{i}\left[F\left(r_{i}(t-1)\right)\right]-\alpha_{i}\left(r_{i}(t-1)\right)\right\} \left|0\right\rangle +\\
 & +\left\{ \beta_{i}\left[F\left(r_{i}(t-1)\right)\right]-\beta_{i}\left(r_{i}(t-1)\right)\right\} \left|1\right\rangle 
\end{aligned}
\end{equation}
from the above equations it follows that the map $F$ reflects the
morphic web connections as well as stochastic factors affecting each
systemic situation $S_{i}$ and, therefore, the corresponding qubit
$\left|\psi_{i}(t)\right\rangle $, which allows for an evolutionary
dynamical framework to be incorporated in the path-dependent quantum
computation.

The interconnected risk dynamics depends strongly upon the coupling
parameter $\varepsilon$ in Eq.(18) to the local mean field $h_{i}(t)$
described by Eq.(20), without this coupling no coevolution would take
place between the different systemic situations. The risk profile
changes in a complex fashion with $\varepsilon$, thus, for instance,
in 1,000 independent Netlogo simulations for $\varepsilon=0.1$, $0.3$,
$0.5$ and $0.7$, with $\theta_{UN}=0.9$ and $d_{UN}=0.95$ (strong
capability of the UN to influence the two countries), $b=4$ and $\delta=1.0E-4$,
we found that for each value of $\varepsilon$, in the corresponding
1,000 simulations, a nuclear war scenario eventually took place (actualization
of a nuclear war) after a few quantum computation steps, which means
that the model captures a dynamics of rising conflituality between
two countries that, given the right conditions, are willing to use
their nuclear arsenal, so that despite the very strong impact of the
UN lowering the threat probability, the two countries still go to
war and for each value of $\varepsilon$, the corresponding 1,000
simulations always led to a nuclear war. The number of quantum computation
steps that took for this to occur, in the simulations, however, differed
critically with $\varepsilon$, as table 1's statistics, presented
in appendix, show.

On average, it took approximately eight quantum computation steps
for these countries to reach nuclear war, however, the maximum number
of steps it takes to reach the nuclear war scenario seems to depend
upon the coupling, thus, for the strong coupling of $\varepsilon=0.7$,
for instance, the maximum is 28 times larger than the second largest
maximum (that occurs for $\varepsilon=0.5$). The minimum, for $\varepsilon=0.7$,
also falls below the previous couplings, with a minimum of 2 quantum
computation steps to reach nuclear war. Half of the simulations have
produced, for each coupling, less than $6$ quantum computation steps
to reach nuclear war (median), and while the most frequent number
of steps is also $6$ for $\varepsilon=0.1$, it becomes $5$ for
the rest of the parameter values.

The simulation statistics, thus, show, in the central tendency statistics,
some structural patterns that do not seem to vary much with the coupling,
the only differences showing up at the extreme values evaluated in
terms of the maxima. Indeed, it is at the extreme values that the
coupling seems to lead to a break in pattern, which also explains
the pattern of rise in heavy tails, especially in the highest coupling
of $\varepsilon=0.7$ with a kurtosis of near 29.

Higher coupling can lead to a higher kurtosis and a higher maximum,
which, in this case, means that higher coupling can lead to some extreme
cases in which it takes much longer for the countries to reach a nuclear
war.

The histograms shown in the figure 1 (presented in appendix) for the
same simulations complete this picture with some additional visible
differences. Indeed, they show that, while as the coupling increases,
in the simulations, there occur cases in which it takes a longer number
of quantum computation steps to reach nuclear war with a greater relative
frequency, the cases in which nuclear war is reached in a smaller
number of steps also take place with a higher relative frequency%
\footnote{This is seen in the first bar of the histograms that rises steadily
with rising coupling, becoming the modal class for $\varepsilon=0.7$.%
}.

Since the model is built from a morphic web that is worked from the
linkages between risk situations that may lead to a nuclear war, it
is not surprising that the model does eventually lead to a nuclear
war scenario, through its path-dependent quantum unitary evolution,
it was built exactly to capture such an extreme risk dynamics scenario,
that is, it is modelling a dynamics of tensions between countries
that possess nuclear weapons and that may use them, given the right
conditions in terms of a very high conflictuality. This feature of
the model may make it effective as a tool for constructing early warning
systems that try to identify risk dynamics between two countries with
nuclear weapons and a history of past political and social tensions,
thus, the model is addressing the potential for regional nuclear conflict
and the consequences that may arise from such conflicts.

\section{Discussion}

Scenario analysis and game theory are two major techniques central
for risk analysis, but while classical game theory has focused on
individuals and populations (in the case of classical evolutionary
game theory), quantum game theory also allows for another type of
application, in which one deals with an \emph{evolutionary scenario
analysis}. In this case, risk mathematics may benefit from the combination
of its \emph{morphic webs}-based formalism, used in the formulation
of general scenarios, with quantum game theory, leading to a scenario
dynamics that follow the structural systemics of the \emph{morphic
webs} of risk situations leading to coevolving dynamics in which one
situation can trigger another.

A statistical analysis of the resulting dynamics, through repeated
experiments, may be helpful in the different applications of risk
mathematics as was shown in the present work, through the application
of this modelling methodology to a geopolitical game with threat of
nuclear war, in which a nuclear war scenario may result from the coevolution
of the political tensions and conflictuality between two countries,
thus addressing, through a quantum game on a \emph{quantum risk structure},
political and social dynamics leading up to regional nuclear conflict,
a methodology that can be expanded to other areas of application of
risk mathematics that demand the employment of \emph{evolutionary
scenario analysis} in the context of interconnected risk situations'
dynamics.

\pagebreak{}

\section*{Appendix}

\begin{table}[H]
\begin{centering}
{\footnotesize }%
\begin{tabular}{|c|c|c|c|c|}
\hline 
{\footnotesize $\varepsilon$} & {\footnotesize 0.1} & {\footnotesize 0.3} & {\footnotesize 0.5} & {\footnotesize 0.7}\tabularnewline
\hline 
{\footnotesize Number of simulations} & {\footnotesize 1,000} & {\footnotesize 1,000} & {\footnotesize 1,000} & {\footnotesize 1,000}\tabularnewline
\hline 
\hline 
{\footnotesize Maximum} & {\footnotesize 49 } & {\footnotesize 43 } & {\footnotesize 52 } & {\footnotesize 80 }\tabularnewline
\hline 
{\footnotesize Minimum} & {\footnotesize 3 } & {\footnotesize 3 } & {\footnotesize 3 } & {\footnotesize 2 }\tabularnewline
\hline 
{\footnotesize Mode} & {\footnotesize 6 } & {\footnotesize 5 } & {\footnotesize 5 } & {\footnotesize 5 }\tabularnewline
\hline 
{\footnotesize Median} & {\footnotesize 6} & {\footnotesize 6} & {\footnotesize 6} & {\footnotesize 6}\tabularnewline
\hline 
{\footnotesize Mean} & {\footnotesize 8.043 } & {\footnotesize 7.911 } & {\footnotesize 8.089 } & {\footnotesize 7.79 }\tabularnewline
\hline 
{\footnotesize Standard-Deviation} & {\footnotesize 5.157} & {\footnotesize 5.586 } & {\footnotesize 6.200 } & {\footnotesize 5.852 }\tabularnewline
\hline 
{\footnotesize Skewness} & {\footnotesize 3.618} & {\footnotesize 2.920 } & {\footnotesize 3.088} & {\footnotesize 3.948}\tabularnewline
\hline 
{\footnotesize Kurtosis} & {\footnotesize 16.107 } & {\footnotesize 10.091 } & {\footnotesize 11.768} & {\footnotesize 28.665 }\tabularnewline
\hline 
\end{tabular}
\par\end{centering}{\footnotesize \par}

\caption{{\footnotesize Statistics from Netlogo experiments for the model,
with parameters: $\theta_{UN}=0.9$, $d_{UN}=0.95$, $b=4$ and $\delta=1.0E-4$,
and $\varepsilon=0.1$, $0.3$, $0.5$ and $0.7$ with 1,000 simulations
for each alternative value of $\varepsilon$. }}
\end{table}

\begin{figure}[H]
\includegraphics[scale=0.3]{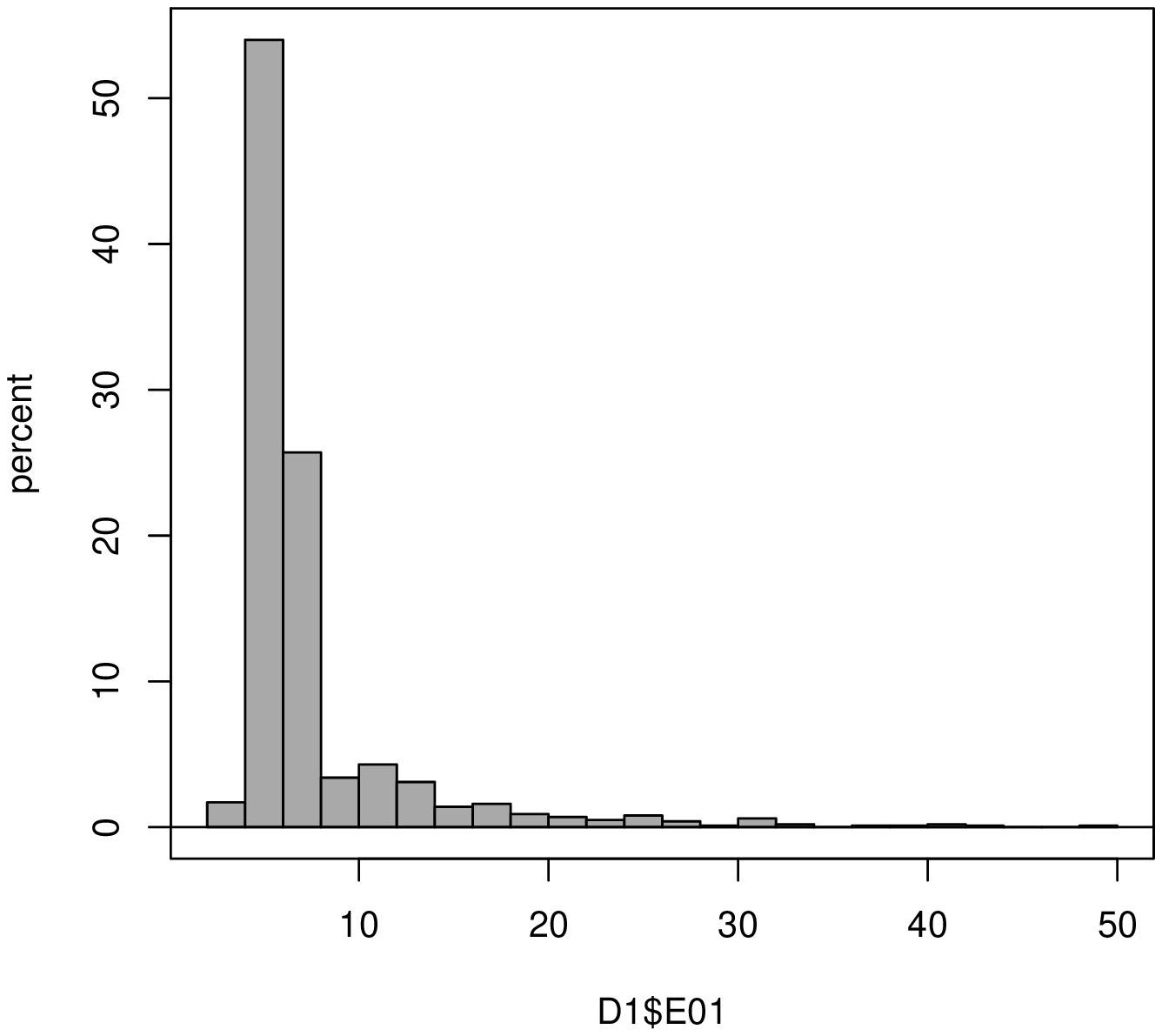}\includegraphics[scale=0.3]{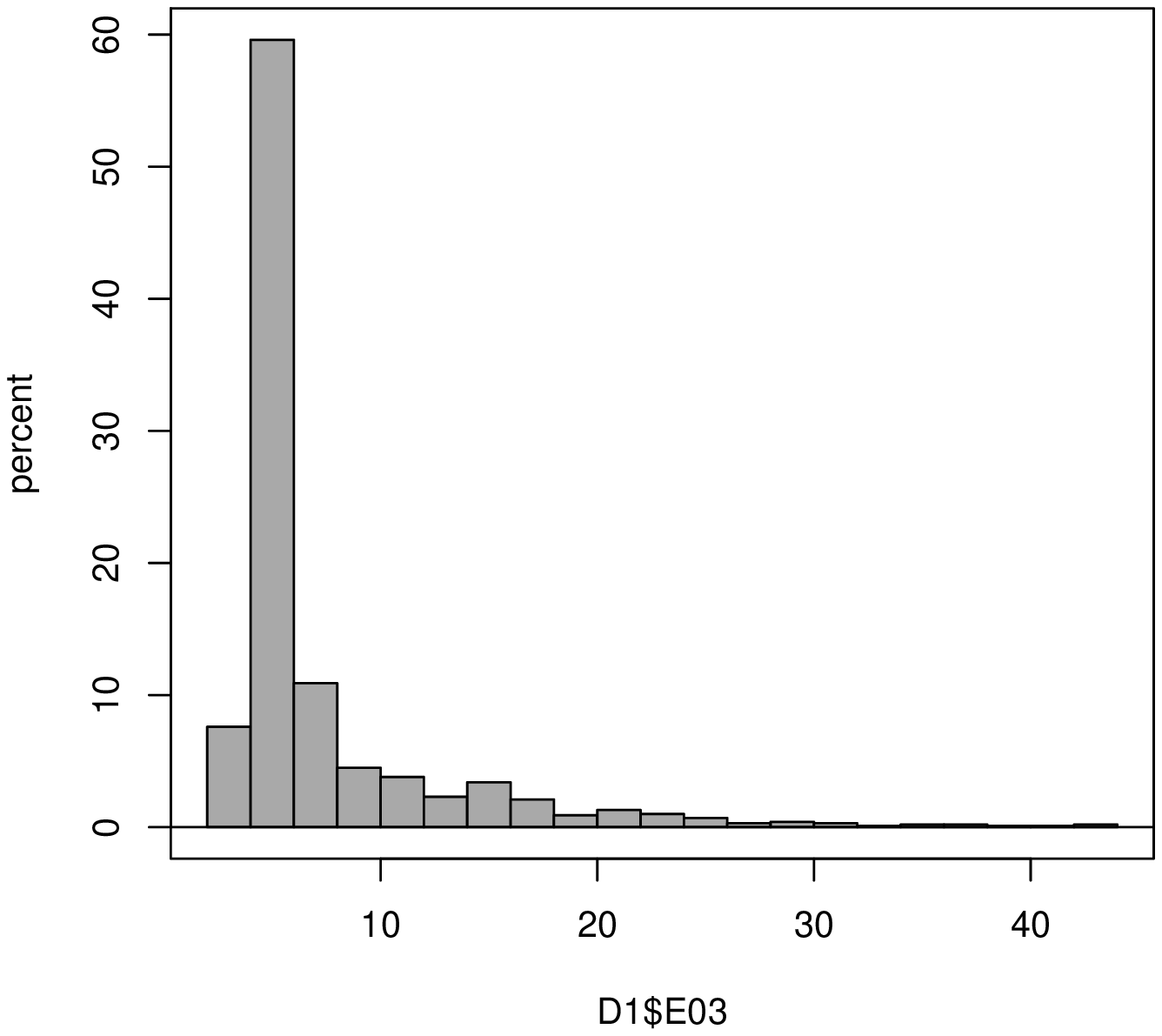}

\includegraphics[scale=0.3]{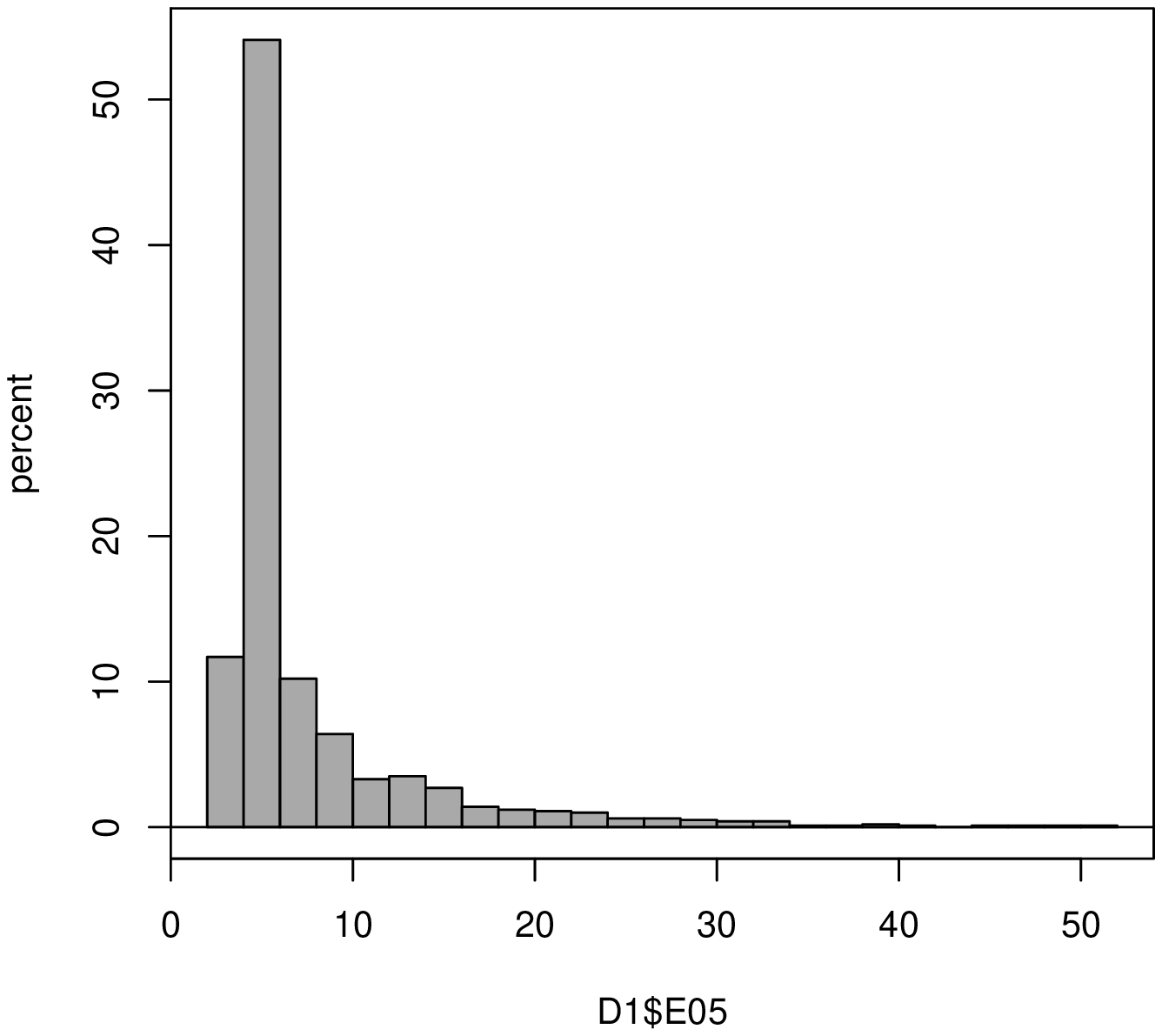}\includegraphics[scale=0.3]{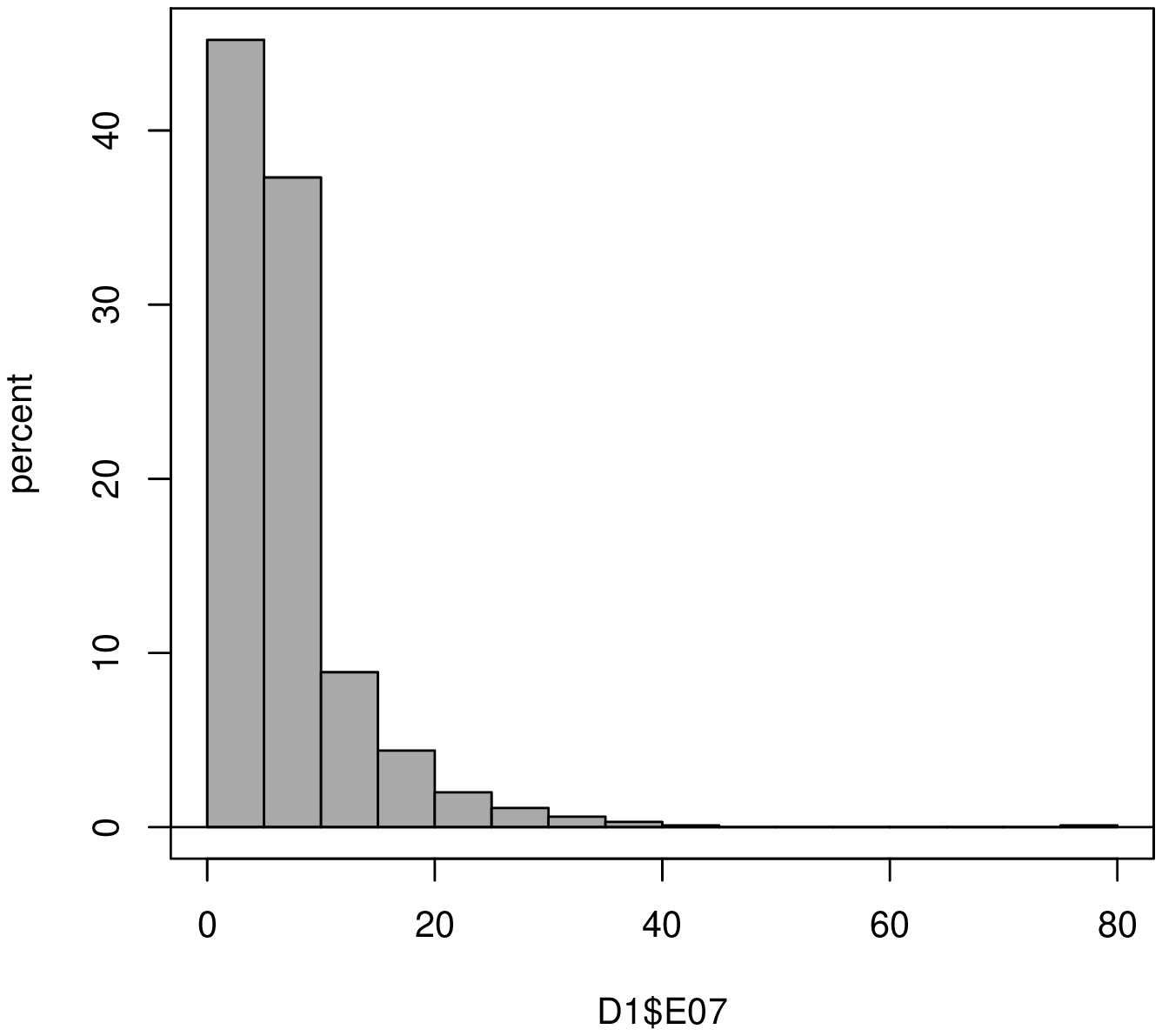}

\caption{{\footnotesize Histograms with 20 bins, estimated for table 1's data.}}
\end{figure}

\end{document}